\documentclass[11pt]{article}
\usepackage[T1]{fontenc}
\usepackage{amsmath,amssymb,mathtools,bm,graphicx}
\usepackage[margin=1in]{geometry}
\usepackage[colorlinks=true,linkcolor=blue,urlcolor=blue]{hyperref}
\newcommand{\thf}{\vartheta_4}
\newcommand{\sn}{\operatorname{sn}}
\newcommand{\cn}{\operatorname{cn}}
\newcommand{\dn}{\operatorname{dn}}
\title{Trigonal Elliptic Coordinates on the Sphere and Some Applications}
\author{Vladimir Mikhalev}
\date{}
\begin{document}
\maketitle
\begin{abstract}
We construct a theta-function parametrization of the unit sphere adapted to the eight trigonal directions. The construction follows from a cubic theta identity discussed by Bobek and admits an equivalent representation in Jacobi elliptic functions. The angular parameter and theta nome give a nonorthogonal coordinate system whose squared Cartesian components are related to the three-component Volterra chain. An algebraic change of angle, retaining the nome, yields orthogonal coordinates with explicit metric coefficients and a diagonal Laplace--Beltrami operator. We apply both parametrizations to the geometry of O'Brien's reduced Jahn--Teller model and give compact expressions for real spherical harmonics of degrees two and three. The orthogonal angle is the polar angle of an $E_g$ quadrupolar doublet, while one octupolar harmonic depends only on the nome within each octant.
\end{abstract}
\section{Theta-function parametrization}
The construction uses three equally shifted Jacobi theta functions. Their normalization maps each choice of Cartesian signs to a spherical octant centered on a trigonal direction. Let $\vartheta_4(z,q)$ denote the fourth Jacobi theta function, with angular argument $z$ and real nome $q$. Define
\begin{equation}
 R(q)^2=2\frac{\thf(2\pi/3,q)}{\thf(0,q)}+\frac{\thf(0,q)^2}{\thf(2\pi/3,q)^2}.
\end{equation}
For independent signs $\varepsilon_i\in\{-1,1\}$, the Cartesian coordinates are
\begin{equation}\label{map}
 \begin{aligned}
 X(t,q)&=\frac{\varepsilon_1}{R(q)}\sqrt{\frac{\thf(t-2\pi/3,q)^2}{\thf(t+2\pi/3,q)\thf(t,q)}},\\
 Y(t,q)&=\frac{\varepsilon_2}{R(q)}\sqrt{\frac{\thf(t+2\pi/3,q)^2}{\thf(t,q)\thf(t-2\pi/3,q)}},\\
 Z(t,q)&=\frac{\varepsilon_3}{R(q)}\sqrt{\frac{\thf(t,q)^2}{\thf(t-2\pi/3,q)\thf(t+2\pi/3,q)}}.
 \end{aligned}
\end{equation}
Positive square roots are used before applying the signs. The parameters range over $t\in\mathbb R/(\pi\mathbb Z)$, $0\leq q<1$.

Each sign choice describes one open spherical octant. At $q=0$, all values of $t$ represent the center $(\varepsilon_1,\varepsilon_2,\varepsilon_3)/\sqrt3$; the map extends continuously to this point, where the angular parameter degenerates. The coordinate great circles $XYZ=0$ form the boundaries of the octants and are excluded from the open charts.

\enlargethispage{3\baselineskip}
\paragraph{Spherical normalization.} The identity $X^2+Y^2+Z^2=1$ follows from Bobek's cubic relation, as shown in Section~2. It can also be verified in Mathematica or established by analyzing the zeros and poles of the theta-function expressions.

\begin{figure}[htbp]
 \centering
 \includegraphics[width=0.95\textwidth,height=0.74\textheight,keepaspectratio]{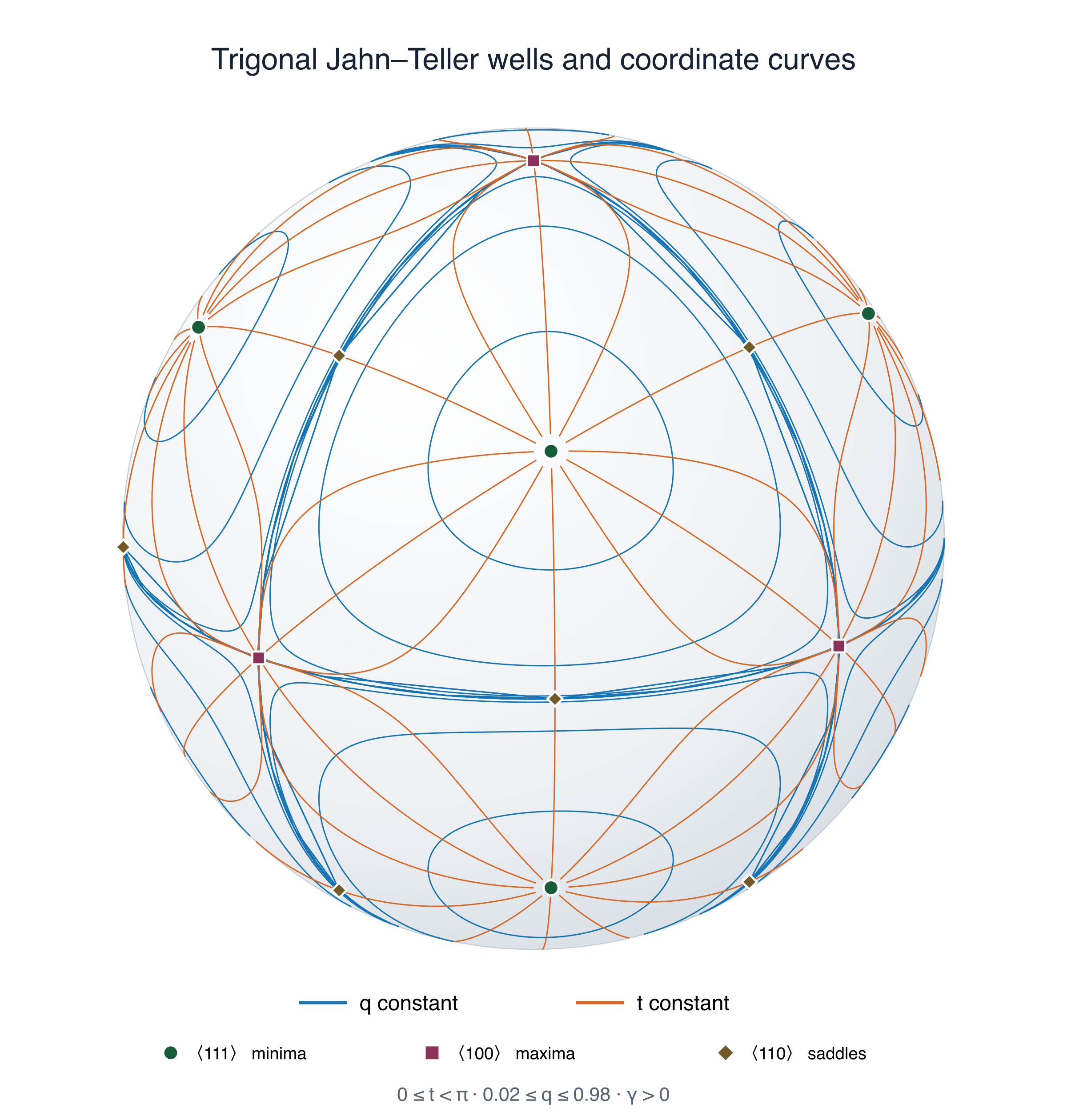}
 \caption{The original theta-function coordinate curves on the sphere. Blue curves have constant $q$; orange curves have constant $t$. In O'Brien's reduced Jahn--Teller model in the strong-coupling adiabatic approximation with positive cubic anisotropy, circles mark trigonal minima, squares tetragonal maxima, and diamonds intermediate saddles. The plotted curves are truncated to $0.02\leq q\leq0.98$; the stationary markers include limiting configurations.}
 \label{fig:original-grid}
\end{figure}
\clearpage

\section{Bobek's cubic identity}
The construction is motivated by Bobek's theta-function parametrization of a plane cubic [\cite{Bobek}, p.~241, Eq.~(16)]. The associated identity [p.~243, Eq.~(18)] is
\begin{equation}
 f_1^3+f_2^3+f_3^3=c f_1f_2f_3.
\end{equation}
Dividing by the nonzero product gives
\begin{equation}
 \frac{f_1^2}{f_2f_3}+\frac{f_2^2}{f_3f_1}+\frac{f_3^2}{f_1f_2}=c.
\end{equation}
On a real branch with $c>0$ and positive ratios, the independently signed square roots of these terms divided by $c$ define Cartesian coordinates on the unit sphere. For the map in Eq.~\eqref{map}, write
\begin{equation}\label{theta-factors}
 A_1=\thf(t-2\pi/3,q),\qquad
 A_2=\thf(t+2\pi/3,q),\qquad
 A_3=\thf(t,q).
\end{equation}
The corresponding normalization identity is
\begin{equation}\label{normalization}
 A_1^3+A_2^3+A_3^3=R(q)^2A_1A_2A_3.
\end{equation}
Dividing by $R(q)^2A_1A_2A_3$ gives $X^2+Y^2+Z^2=1$ directly from Eq.~\eqref{map}.

\subsection{Reduction of Bobek's theta products}
Let $\Omega,\Omega'$ be Bobek's periods and write
\begin{equation}
 \tau_B=\frac{\Omega'}{\Omega},\qquad
 Q=e^{i\pi\tau_B},\qquad q=e^{i\pi\tau_B/3},\qquad
 h=\frac{\pi\tau_B}{3},\qquad z=\frac{\pi u_B}{\Omega}.
\end{equation}
Thus $Q=q^3$: reducing the three-factor product changes Bobek's nome $Q$ to its cube root $q$. In the real parametrization considered here, $0<q<1$, the period ratio is purely imaginary, and $h=-i\log q$.

Choose the standard normalization $\Theta_1(u_B)=\vartheta_1(z,Q)$; any common constant normalization cancels from the ratios. Define
\begin{equation}
 P(z)=\vartheta_1(z,Q)\vartheta_1(z-h,Q)\vartheta_1(z+h,Q).
\end{equation}
Bobek's three functions in Eq.~(16), printed p.~241, are then
\begin{equation}
 f_1=P(z-2\pi/3),\qquad f_2=P(z+2\pi/3),\qquad f_3=P(z).
\end{equation}
The required product reduction is
\begin{equation}\label{product-reduction}
 \boxed{P(z)=\kappa(q)\vartheta_1(z,q),\qquad
 \kappa(q)=-\frac{\vartheta_1'(0,Q)\vartheta_1(h,Q)^2}{\vartheta_1'(0,q)}.}
\end{equation}
Primes here denote differentiation in the first, angular argument.

To prove Eq.~\eqref{product-reduction}, the standard quasiperiodicity formulas~\cite{DLMFtheta} give
\begin{equation}
 P(z+\pi)=-P(z),\qquad
 P(z+h)=-q^{-1}e^{-2iz}P(z).
\end{equation}
These are precisely the multipliers of $\vartheta_1(z,q)$. Both functions have simple zeros on the lattice $\pi\mathbb Z+h\mathbb Z$: the three shifted factors supply the three residue classes of the imaginary-period lattice for nome $Q$. Thus $P(z)/\vartheta_1(z,q)$ extends to an entire doubly periodic function and is constant. Differentiation at $z=0$ gives the stated $\kappa$.

Restrict the complex argument to the line
\begin{equation}
 z=t+\frac h2,\qquad
 \text{equivalently }u_B=\frac{\Omega t}{\pi}+\frac{\Omega'}6.
\end{equation}
The half-period translation formula is
\begin{equation}
 \vartheta_1(t+h/2,q)=i q^{-1/4}e^{-it}\vartheta_4(t,q).
\end{equation}
Consequently, with $\omega=e^{2\pi i/3}$ and
$L(t,q)=i\kappa(q)q^{-1/4}e^{-it}$,
\begin{equation}\label{bobek-to-A}
 \boxed{(f_1,f_2,f_3)=L(t,q)(\omega A_1,\omega^{-1}A_2,A_3).}
\end{equation}
In particular, both the common factor and the phases cancel in the coordinate ratios:
\begin{equation}
 \frac{f_1^2}{f_2f_3}=\frac{A_1^2}{A_2A_3},\qquad
 \frac{f_2^2}{f_3f_1}=\frac{A_2^2}{A_3A_1},\qquad
 \frac{f_3^2}{f_1f_2}=\frac{A_3^2}{A_1A_2}.
\end{equation}
Also $\omega^3=1$, so substitution into Bobek's cubic identity gives
\begin{equation}
 A_1^3+A_2^3+A_3^3=c(Q)A_1A_2A_3.
\end{equation}
At $t=0$, evenness and periodicity give $A_1=A_2=\vartheta_4(2\pi/3,q)$ and $A_3=\vartheta_4(0,q)$. Therefore
\begin{equation}
 c(q^3)=2\frac{\vartheta_4(2\pi/3,q)}{\vartheta_4(0,q)}
 +\frac{\vartheta_4(0,q)^2}{\vartheta_4(2\pi/3,q)^2}=R(q)^2.
\end{equation}
The reduction therefore gives the normalization used in Eq.~\eqref{map}. The endpoint $q=0$ follows by continuity. As an independent check, the product and phase identities were evaluated at nine parameter pairs with 120-digit input precision and agreed to more than 110 decimal digits.

\section{Jacobi elliptic representation}
An equivalent Jacobi elliptic representation makes the trigonometric limit $q\to0$ transparent: $\operatorname{sn}(u\mid m)\to\sin u$, $\operatorname{cn}(u\mid m)\to\cos u$, and $\operatorname{dn}(u\mid m)\to1$ as $m\to0$. The Cartesian coordinates can be written explicitly in this notation as follows.

We use the parameter convention $m=k^2$, as in Mathematica and Abramowitz and Stegun~\cite{AbramowitzStegun1964}. Set
\begin{equation}
 m=\left(\frac{\vartheta_2(0,q)}{\vartheta_3(0,q)}\right)^4,\quad
 K=K(m),\quad u=\frac{2Kt}{\pi},\quad a=\frac{4K}{3},
\end{equation}
\begin{equation}
 s=\sn(a\mid m),\qquad H^2=\frac{1-ms^4}{3-ms^4}.
\end{equation}
Then
\begin{equation}
 \begin{aligned}
 X&=\frac{\varepsilon_1H}{\sqrt{1-ms^2\sn^2(u-a\mid m)}},\\
 Y&=\frac{\varepsilon_2H}{\sqrt{1-ms^2\sn^2(u+a\mid m)}},\\
 Z&=\frac{\varepsilon_3H}{\sqrt{1-ms^2\sn^2(u\mid m)}}.
 \end{aligned}
\end{equation}
High-precision evaluations of the theta and Jacobi representations agreed to approximately 120 decimal digits at the sampled points.

\section{Relation to the Volterra lattice}
The Volterra lattice, also known as the Kac--van Moerbeke lattice~\cite{KacVanMoerbeke1975}, is the differential--difference system
\begin{equation}\label{volterra-definition}
 \frac{du_n}{d\tau}=u_n(u_{n+1}-u_{n-1}).
\end{equation}
We use its three-periodic reduction $u_{n+3}=u_n$. The exponential variables in Ref.~\cite{KacVanMoerbeke1975} give this quadratic form under $u_n=e^{-R_n}$.
The squares $X^2,Y^2,Z^2$ satisfy the standard periodic three-component Volterra chain~\cite{KacVanMoerbeke1975} after a rescaling of time. The coordinates themselves satisfy the cubic system
\begin{equation}\label{cartesian-volterra}
 \begin{aligned}
 \partial_t X&=\frac{\lambda(q)}2 X(Y^2-Z^2),\\
 \partial_t Y&=\frac{\lambda(q)}2 Y(Z^2-X^2),\\
 \partial_t Z&=\frac{\lambda(q)}2 Z(X^2-Y^2).
 \end{aligned}
\end{equation}
The same equations apply in each octant with its fixed Cartesian signs. Setting $U_1=X^2,U_2=Y^2,U_3=Z^2$ gives
\begin{equation}\label{volterra}
 \partial_tU_i=\lambda(q)U_i(U_{i+1}-U_{i-1}),\qquad i\pmod3,
\end{equation}
where
\begin{equation}
 \lambda(q)=-\frac{4K}{\pi H^2}\frac{\cn(a\mid m)\dn(a\mid m)}{\sn(a\mid m)}.
\end{equation}
Thus $\tau=\lambda(q)t$ gives the periodic three-component Volterra chain. Its invariants satisfy
\begin{equation}
 \sum_iU_i=1,\qquad \prod_iU_i=R(q)^{-6}.
\end{equation}
The relation in Eq.~\eqref{volterra} was identified using Mathematica. Symbolic differentiation followed by high-precision numerical evaluation at 30 parameter pairs gave residuals vanishing to approximately 139 decimal digits. This computational evidence is the basis for the Volterra connection used here; a separate analytic proof is outside the scope of the present work.

\section{Application to a dynamic Jahn--Teller model}
O'Brien's reduced Jahn--Teller model~\cite{OBrien} describes an orbital triplet coupled to both $E_g$ and $T_{2g}$ vibrations in a strong-coupling adiabatic approximation. The reduced nuclear motion takes place on a two-dimensional distortion surface represented by a sphere, with angular Hamiltonian
\begin{equation}
 \widehat{\mathcal H}=-B_{\rm rot}\Delta_{S^2}
 +\gamma\left(X^4+Y^4+Z^4-\frac35\right),\qquad
 B_{\rm rot}=\frac{\hbar^2\omega^2}{6\rho^2}.
\end{equation}
Here $\rho$ denotes the fixed radial distortion amplitude called $q$ in O'Brien's paper; it is distinct from our theta nome.

\subsection{Five vibrational coordinates}
For O'Brien's reduced Jahn--Teller model, Eq.~(2.7) of Ref.~\cite{OBrien} becomes
\begin{equation}
 \begin{aligned}
 q_\theta&=\frac\rho2(3Z^2-1),&
 q_\epsilon&=\frac{\sqrt3\rho}{2}(X^2-Y^2),\\
 q_\xi&=\sqrt3\rho YZ,&q_\eta&=\sqrt3\rho ZX,&
 q_\zeta&=\sqrt3\rho XY.
 \end{aligned}
\end{equation}
Substituting Eq.~\eqref{map} gives an explicit theta-function representation of all five scaled vibrational coordinates. Their squared sum is $\rho^2$. At the $q=0$ centers,
\begin{equation}
 q_\theta=q_\epsilon=0,\qquad
 (q_\xi,q_\eta,q_\zeta)=\frac\rho{\sqrt3}
 (\varepsilon_2\varepsilon_3,\varepsilon_3\varepsilon_1,\varepsilon_1\varepsilon_2).
\end{equation}
Opposite sphere points give the same distortion, yielding four physical trigonal configurations.

\subsection{Electronic-basis rotation}
The electronic-basis rotation in O'Brien's reduced Jahn--Teller model is given by Eq.~(2.12) of Ref.~\cite{OBrien}. Writing $s_\perp=\sqrt{X^2+Y^2}$, it takes the form
\begin{equation}
 \begin{pmatrix}|\xi'\rangle\\|\eta'\rangle\\|\zeta'\rangle\end{pmatrix}
 =\begin{pmatrix}
 ZX/s_\perp&ZY/s_\perp&-s_\perp\\
 -Y/s_\perp&X/s_\perp&0\\
 X&Y&Z
 \end{pmatrix}
 \begin{pmatrix}|\xi\rangle\\|\eta\rangle\\|\zeta\rangle\end{pmatrix}.
\end{equation}
The rotation is regular throughout every open octant, including its center at $q=0$. Its first two rows retain O'Brien's transverse-frame convention; they need not coincide with the normalized tangents of the new coordinates.

\subsection{Potential geometry}
The dimensionless potential in the original parameters is
\begin{equation}
 \begin{aligned}
 F(t,q)=\frac{V}{\gamma}=\frac1{R(q)^4}\biggl[&
 \frac{\thf(t-2\pi/3,q)^4}{\thf(t+2\pi/3,q)^2\thf(t,q)^2}\\
 &+\frac{\thf(t+2\pi/3,q)^4}{\thf(t,q)^2\thf(t-2\pi/3,q)^2}\\
 &+\frac{\thf(t,q)^4}{\thf(t-2\pi/3,q)^2\thf(t+2\pi/3,q)^2}\biggr]-\frac35.
 \end{aligned}
\end{equation}
Its stationary values are
\begin{equation}
 F_{\langle111\rangle}=-\frac4{15},\qquad
 F_{\langle110\rangle}=-\frac1{10},\qquad
 F_{\langle100\rangle}=\frac25.
\end{equation}
For $\gamma>0$, trigonal directions are minima and tetragonal directions are maxima; their roles reverse for $\gamma<0$. Intermediate directions remain saddles.

For $\gamma<0$, polar coordinates centered on a selected tetragonal axis are adapted to local motion about the corresponding well. For $\gamma>0$, each trigonal minimum is a $q=0$ center of the present parametrization. The coordinates are therefore adapted to zero-point motion localized near these minima when the anisotropy is sufficiently strong relative to the angular kinetic scale. The application considered here concerns this geometry and its explicit coordinate representation.

\begin{figure}[htbp]
 \centering
 \includegraphics[width=\textwidth]{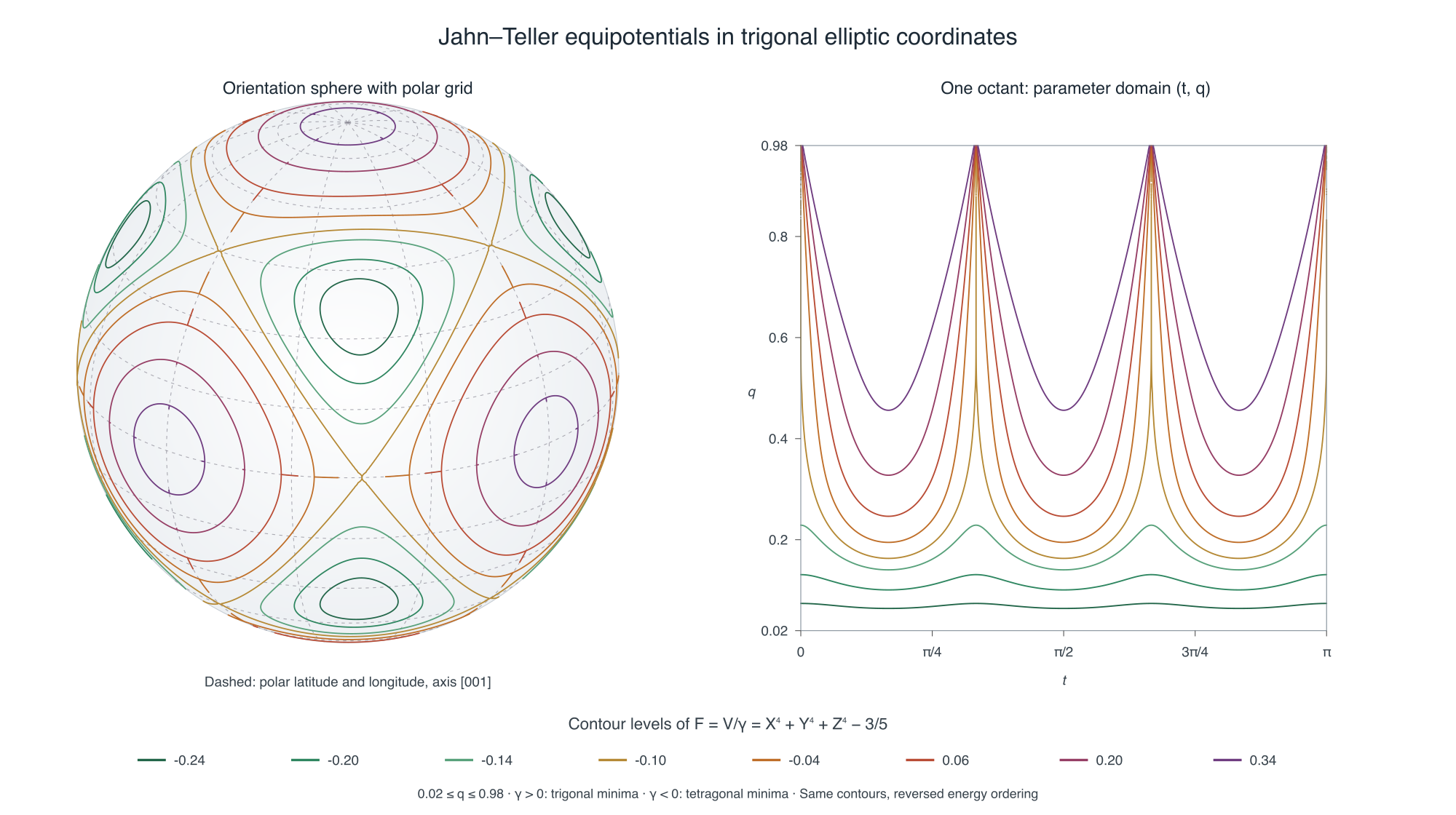}
 \caption{Equipotentials evaluated using the explicit $(t,q)$ parametrization. Left: sphere with a dashed polar grid aligned with $[001]$. Right: the same levels in one octant's parameter domain. Reversing the sign of $\gamma$ reverses energy ordering but preserves contour geometry. The saddle level is $F=-0.10$. Endpoint neighborhoods are omitted: $0.02\leq q\leq0.98$.}
\end{figure}

\subsection{Explicit metric and kinetic operator}
Use the theta-argument shifts $(\beta_1,\beta_2,\beta_3)=(-2\pi/3,2\pi/3,0)$ and define
\begin{equation}
 \begin{aligned}
 p_i&=\frac32\,\partial_t\log\thf(t+\beta_i,q)
      -\frac12\sum_{j=1}^3\partial_t\log\thf(t+\beta_j,q),\\
 s_i&=-\frac{3}{8q}\frac{\partial_t^2\thf(t+\beta_i,q)}{\thf(t+\beta_i,q)}
      +\frac{1}{8q}\sum_{j=1}^3\frac{\partial_t^2\thf(t+\beta_j,q)}{\thf(t+\beta_j,q)}
      -\frac{R'}R.
 \end{aligned}
\end{equation}
The expression for $s_i$ uses the theta heat equation $4q\,\partial_q\thf(t+\beta_i,q)=-\partial_t^2\thf(t+\beta_i,q)$ for $q>0$. Here the $q$-derivative is taken at fixed $t$ in the original $(t,q)$ parametrization, rather than at fixed $\alpha$ as in Section~6.
Then logarithmic differentiation gives
\begin{equation}
 \begin{aligned}
 (X_t,Y_t,Z_t)&=(Xp_1,Yp_2,Zp_3),\\
 (X_q,Y_q,Z_q)&=(Xs_1,Ys_2,Zs_3),
 \end{aligned}
\end{equation}
giving the explicit metric
\begin{equation}
 \boxed{\begin{aligned}
 E&=X^2p_1^2+Y^2p_2^2+Z^2p_3^2,\\
 F&=X^2p_1s_1+Y^2p_2s_2+Z^2p_3s_3,\\
 G&=X^2s_1^2+Y^2s_2^2+Z^2s_3^2.
 \end{aligned}}
\end{equation}
The area density has a simple form. Define
\begin{equation}
 C(q)=X^2Y^2Z^2=R^{-6},\qquad P=X^2Y^2+Y^2Z^2+Z^2X^2.
\end{equation}
The Volterra relation gives $E=\lambda^2(P-9C)/4$, while direct differentiation of the invariant on the sphere gives
\begin{equation}
 \|\nabla_{S^2}C\|^2=4C(P-9C).
\end{equation}
Since $C_t=0$, $C_q=C'$, the component of $(X_q,Y_q,Z_q)$ normal to a fixed-$q$ curve has squared length $C'^2/\|\nabla_{S^2}C\|^2$. Hence
\begin{equation}
 \boxed{J^2=EG-F^2=\frac{\lambda^2C'^2}{16C},\qquad
 J=\frac{3|\lambda R'|}{2R^4}.}
\end{equation}
Thus the area density $J$ depends only on $q$. The expressions obtained from the Volterra relation agree with the direct theta-derivative metric to more than 110 digits at the sampled points. The Volterra relation is used with the computational verification described in Section~4.

The metric inverse is $g^{-1}=J^{-2}\left(\begin{smallmatrix}G&-F\\-F&E\end{smallmatrix}\right)$, so the kinetic operator is
\begin{equation}\label{kinetic}
 \boxed{\widehat T\psi=-\frac{B_{\rm rot}}J\left\{
 \partial_t\left(\frac{G\psi_t-F\psi_q}{J}\right)
 +\partial_q\left(\frac{E\psi_q-F\psi_t}{J}\right)\right\}.}
\end{equation}
Equivalently, since $J=J(q)$,
\begin{equation}
 \begin{aligned}
 \widehat T\psi=-\frac{B_{\rm rot}}{J^2}\biggl[&G\psi_{tt}-2F\psi_{tq}+E\psi_{qq}\\
 &+\left(G_t-F_q+F\frac{J'}J\right)\psi_t
 +\left(E_q-F_t-E\frac{J'}J\right)\psi_q\biggr].
 \end{aligned}
\end{equation}
The unit-sphere area measure is $J\,dt\,dq$. These expressions apply on regular coordinate patches; smoothness at $q=0$ must be imposed through the limiting behavior at the octant center.

Section~6 constructs an orthogonal reparametrization that retains the nome.

The adiabatic electronic state is $X|\xi\rangle+Y|\eta\rangle+Z|\zeta\rangle$. Since antipodal sphere points represent the same distortion and the electronic state changes sign under inversion, the nuclear amplitude must satisfy $\psi(-X,-Y,-Z)=-\psi(X,Y,Z)$ for the complete vibronic state to be single-valued. The amplitudes must also match smoothly across octant boundaries and be regular at the centers.

\section{Orthogonal reparametrization}
Retaining $q$, an orthogonal angular parameter must satisfy $t_q|_\alpha=-F/E$. This equation admits an algebraic solution. In terms of the original theta functions, define
\begin{equation}\label{alpha-map}
 \boxed{\tan\alpha=
 \frac{\sqrt3\bigl[\thf(t+2\pi/3,q)^3-\thf(t,q)^3\bigr]}
 {2\thf(t-2\pi/3,q)^3-\thf(t+2\pi/3,q)^3-\thf(t,q)^3}.}
\end{equation}
The quadrant of $\alpha\in\mathbb R/(2\pi\mathbb Z)$ is determined by taking the signs of the numerator and denominator as the signs of $\sin\alpha$ and $\cos\alpha$, respectively. The common positive normalization cancels in this ratio. Define
\begin{equation}
 \begin{aligned}
 \sigma^2=1-\frac{3}{R(q)^4}\biggl[&
 \frac{\thf(t-2\pi/3,q)\thf(t+2\pi/3,q)}{\thf(t,q)^2}\\
 &+\frac{\thf(t+2\pi/3,q)\thf(t,q)}{\thf(t-2\pi/3,q)^2}
 +\frac{\thf(t,q)\thf(t-2\pi/3,q)}{\thf(t+2\pi/3,q)^2}\biggr],
 \qquad \sigma\geq0.
 \end{aligned}
\end{equation}
Then the explicit theta expressions obey
\begin{equation}\label{new-U}
 \begin{aligned}
 \frac{\thf(t-2\pi/3,q)^2}{R(q)^2\thf(t+2\pi/3,q)\thf(t,q)}
 &=\frac{1+2\sigma\cos\alpha}{3},\\
 \frac{\thf(t+2\pi/3,q)^2}{R(q)^2\thf(t,q)\thf(t-2\pi/3,q)}
 &=\frac{1+2\sigma\cos(\alpha-2\pi/3)}{3},\\
 \frac{\thf(t,q)^2}{R(q)^2\thf(t-2\pi/3,q)\thf(t+2\pi/3,q)}
 &=\frac{1+2\sigma\cos(\alpha+2\pi/3)}{3}.
 \end{aligned}
\end{equation}
The angle $\alpha$ is measured from the center $(1/3,1/3,1/3)$ in the plane of squared Cartesian coordinates. Below, the angular shifts are $\delta_1=0$, $\delta_2=-2\pi/3$, and $\delta_3=2\pi/3$.

\subsection{Algebraic map and admissible root}
With $C(q)=R(q)^{-6}$, multiplication of the three identities in Eq.~\eqref{new-U} gives
\begin{equation}\label{sigma-cubic}
 \boxed{1-3\sigma^2+2\sigma^3\cos3\alpha=27C(q).}
\end{equation}
For each $\alpha$ and $0<q<1$, select the unique root
\begin{equation}
 0<\sigma<\sigma_b(\alpha),\qquad
 \sigma_b=-\frac1{2\min_i\cos(\alpha+\delta_i)}.
\end{equation}
This interval ensures $X^2,Y^2,Z^2>0$. The left side of Eq.~\eqref{sigma-cubic} decreases from 1 to 0 on that interval; its derivative is $-6\sigma(1-\sigma\cos3\alpha)<0$ in the interior. Therefore the new sphere map is completely specified by
\begin{equation}
 \boxed{\begin{aligned}
 X(\alpha,q)&=\varepsilon_1\sqrt{\frac{1+2\sigma(\alpha,q)\cos\alpha}{3}},\\
 Y(\alpha,q)&=\varepsilon_2\sqrt{\frac{1+2\sigma(\alpha,q)\cos(\alpha-2\pi/3)}{3}},\\
 Z(\alpha,q)&=\varepsilon_3\sqrt{\frac{1+2\sigma(\alpha,q)\cos(\alpha+2\pi/3)}{3}}.
 \end{aligned}}
\end{equation}
The selected cubic root specifies the map algebraically. At $q=0$, $\sigma=0$ and all values of $\alpha$ represent the octant center.

\subsection{Analytic proof of orthogonality}
Write $D=1-\sigma\cos3\alpha$. Implicit differentiation gives
\begin{equation}
 \sigma_q=-\frac{9C'}{2\sigma D},\qquad
 \sigma_\alpha=-\frac{\sigma^2\sin3\alpha}{D}.
\end{equation}
At fixed $\alpha$, differentiating the algebraic map gives
\begin{equation}
 \partial_q(X^2,Y^2,Z^2)=\frac{\sigma_q}{\sigma}
 \left(X^2-\frac13,Y^2-\frac13,Z^2-\frac13\right).
\end{equation}
At fixed $q$, both $X^2+Y^2+Z^2=1$ and $X^2Y^2Z^2=C(q)$ are constant. Consequently,
\begin{equation}
 \partial_\alpha(X^2+Y^2+Z^2)=0,\qquad
 \partial_\alpha\log(X^2Y^2Z^2)=0.
\end{equation}
It follows that
\begin{equation}
 \begin{aligned}
 X_\alpha X_q+Y_\alpha Y_q+Z_\alpha Z_q
 =\frac{\sigma_q}{4\sigma}\biggl[&\partial_\alpha(X^2+Y^2+Z^2)\\
 &-\frac13\partial_\alpha\log(X^2Y^2Z^2)\biggr]=0.
 \end{aligned}
\end{equation}
The orthogonality proof uses only the algebraic map and its two invariants.

\subsection{Diagonal metric and kinetic operator}
Direct substitution into $g_{ab}=X_aX_b+Y_aY_b+Z_aZ_b$ yields
\begin{equation}\label{orth-metric}
 \boxed{h_\alpha^2=\frac{\sigma^2}{2D},\qquad
 h_q^2=\frac{3C'^2}{8C\sigma^2D},\qquad g_{\alpha q}=0.}
\end{equation}
Define the transformed area density and a scalar factor by
\begin{equation}
 W(q)=\frac{\sqrt3|C'|}{2\sqrt C},\qquad
 \mathcal J=h_\alpha h_q=\frac{W(q)}{2D}.
\end{equation}
The transformed density $\mathcal J$ generally depends on both $\alpha$ and $q$. The kinetic operator becomes
\begin{equation}\label{orth-kinetic}
 \boxed{\widehat T\psi=-\frac{2B_{\rm rot}D}{W}
 \left[\partial_\alpha\left(\frac{W}{\sigma^2}\partial_\alpha\psi\right)
 +\partial_q\left(\frac{\sigma^2}{W}\partial_q\psi\right)\right].}
\end{equation}
The area measure is $\mathcal J\,d\alpha\,dq$. The coefficients are specified by the admissible cubic root and $C(q)$, and the operator contains no mixed derivative. Near $q=0$, $\sigma=3q+O(q^2)$, and to leading order
\begin{equation}
 ds^2\sim\frac92\left(dq^2+q^2d\alpha^2\right),
\end{equation}
showing the usual polar-type degeneracy at the center.

\subsection{Magnified comparison near a trigonal center}
To compare the two angular families, match their labels at the center by
$\alpha_0(t)=\pi/3-2t$. Expansion of the explicit theta formula gives
\begin{equation}
 \alpha(t,q)=\frac\pi3-2t-q\sin6t+O(q^2)
\end{equation}
on a continuous angular branch. The two angular families therefore differ continuously from their common limit as $q$ increases. Their constant-$q$ curves coincide, while the crossing angular curves generally differ. Along the symmetry directions $\sin6t=0$, the matched angular curves coincide.

A useful measure of the deviation is the wrapped angular-label difference
\begin{equation}
 d(q)=\max_{0\leq t<\pi}
 \left|\operatorname{Arg}\exp\!\left(i[\alpha(t,q)-\alpha_0(t)]\right)\right|.
\end{equation}
It satisfies $d(q)=q+O(q^2)$ in radians. Direct evaluation of the theta formulas gives $d=1^\circ$ at $q\simeq0.01745$ and $d=5^\circ$ at $q\simeq0.08716$. These values quantify chosen angular-label tolerances; they do not mark singularities or physical transitions. For example, $d(0.05)\simeq2.866^\circ$ and $d(0.10)\simeq5.739^\circ$.

\begin{figure}[htbp]
 \centering
 \includegraphics[width=\textwidth]{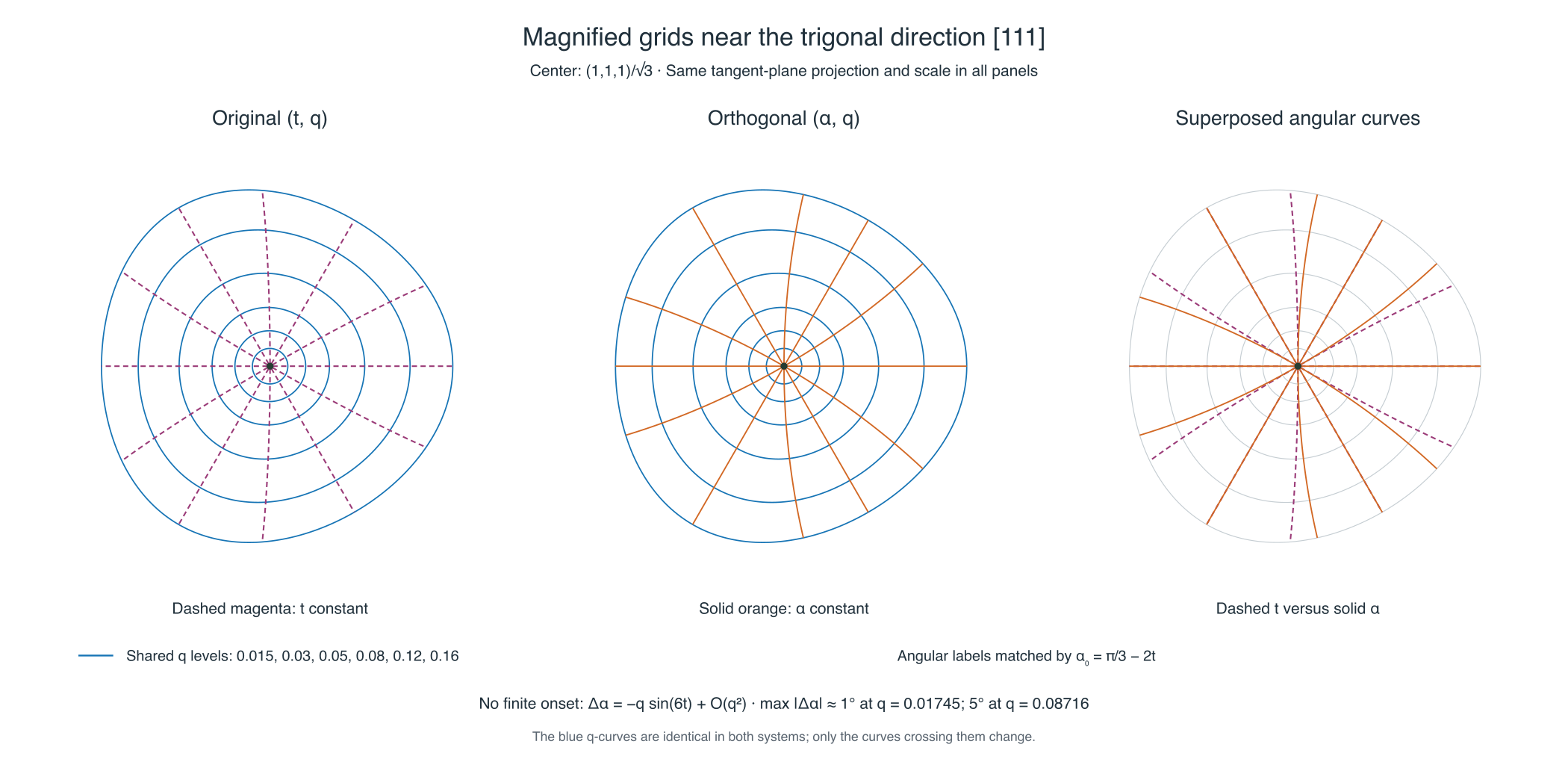}
 \caption{Magnified comparison near $(1,1,1)/\sqrt3$, with identical orthographic tangent-plane projections and scales. Left: original $t,q$ grid; center: orthogonal $\alpha,q$ grid; right: the angular families superposed with labels matched by $\alpha_0=\pi/3-2t$. Constant-$q$ curves are shared, at $q=0.015,0.03,0.05,0.08,0.12,0.16$. Dashed magenta: $t$ constant; solid orange: $\alpha$ constant, at multiples of $\pi/6$. The projection uses tangent directions $(2,-1,-1)/\sqrt6$ and $(0,1,-1)/\sqrt2$.}
 \label{fig:local-grids}
\end{figure}

\subsection{Relation to the original angle and potential}
The coordinate transformation has mixed coefficient $t_\alpha(F+E t_q)$. Its vanishing gives $t_q|_\alpha=-F/E$. Using the Volterra relation fixes the angular orientation:
\begin{equation}
 \alpha_t=-\frac\lambda{\sqrt3}D,\qquad
 t_\alpha=-\frac{\sqrt3}{\lambda D}.
\end{equation}
With $\alpha=\pi/3$ at $t=0$, the unwrapped inverse is
\begin{equation}
 t(\alpha,q)=\frac{\sqrt3}{\lambda(q)}
 \int_\alpha^{\pi/3}\frac{d\beta}{1-\sigma(\beta,q)\cos3\beta}
 \pmod\pi.
\end{equation}
As $t$ increases by $\pi$, $\alpha$ decreases by $2\pi$. The quadrature recovers the original parameter; the orthogonal map itself is evaluated directly from the cubic root.

Finally, $X^4+Y^4+Z^4=(1+2\sigma^2)/3$, so the potential simplifies to
\begin{equation}
 \boxed{V(\alpha,q)=\gamma\left(\frac23\sigma(\alpha,q)^2-\frac4{15}\right).}
\end{equation}
Equipotentials are exactly constant-$\sigma$ curves, not generally constant-$q$ curves. The antipodal condition is unchanged: $\psi_{-\bm\varepsilon}(\alpha,q)=-\psi_{\bm\varepsilon}(\alpha,q)$.

\paragraph{Verification.}
At nine parameter pairs, transforming the original theta-derivative metric with Eq.~\eqref{alpha-map} gave mixed coefficients zero to at least approximately 90 decimal digits and agreement with Eq.~\eqref{orth-metric} to at least approximately 87 digits, using 100-digit inputs.

\section{Spherical harmonics in the two parametrizations}\label{sec:harmonics}
The coordinate maps provide explicit representations of the same spherical harmonics in $(t,q)$ and $(\alpha,q)$. For the cubic symmetry of the Jahn--Teller model, a real basis adapted to the cubic group is particularly useful. The basis functions are real linear combinations of the conventional complex spherical harmonics $Y_{\ell m}$~\cite{AbramowitzStegun1964,DLMFharmonics}. We normalize them on the full unit sphere with measure $d\Omega$.

\subsection{Quadrupolar harmonics: \texorpdfstring{$\ell=2$}{l=2}}
Choose the real basis
\begin{equation}\label{quadrupole-basis}
 \begin{aligned}
 d_1&=\frac{3X^2-1}{2},&
 d_2&=\frac{\sqrt3}{2}(Y^2-Z^2),\\
 d_3&=\sqrt3\,YZ,&d_4&=\sqrt3\,ZX,&d_5&=\sqrt3\,XY.
 \end{aligned}
\end{equation}
The first pair spans the $E_g$ subspace and the last three span $T_{2g}$. This $E_g$ pair is related to the pair used in O'Brien's reduced Jahn--Teller model by an orthogonal change of basis within that two-dimensional space. All five functions have squared norm $4\pi/5$, so
\begin{equation}
 \mathcal Y_{2a}=\sqrt{\frac5{4\pi}}\,d_a,\qquad a=1,\ldots,5,
\end{equation}
form an orthonormal basis.

In the orthogonal coordinates, the $E_g$ pair takes the particularly simple form
\begin{equation}\label{quadrupole-angle}
 \boxed{d_1=\sigma(\alpha,q)\cos\alpha,\qquad
 d_2=\sigma(\alpha,q)\sin\alpha.}
\end{equation}
Thus $\alpha$ is exactly the polar angle of this quadrupolar doublet, and $\sigma$ is its amplitude. This provides a direct symmetry interpretation of the angle introduced in Section~6.

The remaining components are
\begin{equation}\label{quadrupole-offdiag}
 \boxed{\begin{aligned}
 d_3&=\frac{\varepsilon_2\varepsilon_3}{\sqrt3}
 \sqrt{\bigl[1+2\sigma\cos(\alpha-2\pi/3)\bigr]
       \bigl[1+2\sigma\cos(\alpha+2\pi/3)\bigr]},\\
 d_4&=\frac{\varepsilon_3\varepsilon_1}{\sqrt3}
 \sqrt{\bigl[1+2\sigma\cos(\alpha+2\pi/3)\bigr]
       \bigl[1+2\sigma\cos\alpha\bigr]},\\
 d_5&=\frac{\varepsilon_1\varepsilon_2}{\sqrt3}
 \sqrt{\bigl[1+2\sigma\cos\alpha\bigr]
       \bigl[1+2\sigma\cos(\alpha-2\pi/3)\bigr]}.
 \end{aligned}}
\end{equation}
Here $\sigma=\sigma(\alpha,q)$ is the physical root defined in Section~6.

In the original $(t,q)$ parametrization, all five components are explicitly
\begin{equation}\label{quadrupole-theta}
 \begin{aligned}
 d_1&=\frac12\left[
 \frac{3\vartheta_4(t-2\pi/3,q)^2}
 {R(q)^2\vartheta_4(t+2\pi/3,q)\vartheta_4(t,q)}-1\right],\\[3pt]
 d_2&=\frac{\sqrt3}{2R(q)^2}\left[
 \frac{\vartheta_4(t+2\pi/3,q)^2}{\vartheta_4(t,q)\vartheta_4(t-2\pi/3,q)}
 -\frac{\vartheta_4(t,q)^2}{\vartheta_4(t-2\pi/3,q)\vartheta_4(t+2\pi/3,q)}\right],\\[3pt]
 d_3&=\frac{\sqrt3\,\varepsilon_2\varepsilon_3}{R(q)^2}
 \frac{\sqrt{\vartheta_4(t+2\pi/3,q)\vartheta_4(t,q)}}{\vartheta_4(t-2\pi/3,q)},\\[3pt]
 d_4&=\frac{\sqrt3\,\varepsilon_3\varepsilon_1}{R(q)^2}
 \frac{\sqrt{\vartheta_4(t,q)\vartheta_4(t-2\pi/3,q)}}{\vartheta_4(t+2\pi/3,q)},\\[3pt]
 d_5&=\frac{\sqrt3\,\varepsilon_1\varepsilon_2}{R(q)^2}
 \frac{\sqrt{\vartheta_4(t-2\pi/3,q)\vartheta_4(t+2\pi/3,q)}}{\vartheta_4(t,q)}.
 \end{aligned}
\end{equation}
These quadrupolar functions give the same five-dimensional harmonic space as the vibrational-coordinate map in Section~5.1.

\subsection{Octupolar harmonics: \texorpdfstring{$\ell=3$}{l=3}}
A complete real basis consists of a singlet and two triplets:
\begin{equation}\label{octupole-basis}
 \begin{aligned}
 f_0&=XYZ,\\
 (f_1^{(1)},f_2^{(1)},f_3^{(1)})&=
 \bigl(X(5X^2-3),Y(5Y^2-3),Z(5Z^2-3)\bigr),\\
 (f_1^{(2)},f_2^{(2)},f_3^{(2)})&=
 \bigl(X(Y^2-Z^2),Y(Z^2-X^2),Z(X^2-Y^2)\bigr).
 \end{aligned}
\end{equation}
Under the cubic group these span $A_{2u}$, $T_{1u}$, and $T_{2u}$, respectively. Their normalized versions are
\begin{equation}
 \mathcal Y_{30}=\sqrt{\frac{105}{4\pi}}\,f_0,\qquad
 \mathcal Y_{3i}^{(1)}=\sqrt{\frac7{16\pi}}\,f_i^{(1)},\qquad
 \mathcal Y_{3i}^{(2)}=\sqrt{\frac{105}{16\pi}}\,f_i^{(2)}.
\end{equation}
The index $0$ here labels the cubic singlet; it does not denote the usual magnetic quantum number $m=0$.

The singlet has the same simple form in both parametrizations:
\begin{equation}\label{xyz-harmonic}
 \boxed{f_0=XYZ=\frac{\varepsilon_1\varepsilon_2\varepsilon_3}{R(q)^3}.}
\end{equation}
Within a fixed octant, the singlet depends only on $q$. Its Cartesian sign factor ensures the odd inversion parity of the global $\ell=3$ function.

A second simplification follows directly from the coordinate evolution equations in Section~4:
\begin{equation}\label{harmonic-volterra}
 \boxed{(f_1^{(2)},f_2^{(2)},f_3^{(2)})=\frac2{\lambda(q)}(X_t,Y_t,Z_t).}
\end{equation}
The $T_{2u}$ octupolar triplet is therefore proportional to the derivatives of the $\ell=1$ Cartesian functions along the original Volterra evolution. This identity is specific to the coordinate flow described in Section~4.

In the orthogonal parametrization, the two triplets are explicitly
\begin{equation}\label{octupole-orthogonal}
 \boxed{\begin{aligned}
 f_1^{(1)}&=\frac{\varepsilon_1}{3\sqrt3}\sqrt{1+2\sigma\cos\alpha}
 \bigl(10\sigma\cos\alpha-4\bigr),\\
 f_2^{(1)}&=\frac{\varepsilon_2}{3\sqrt3}\sqrt{1+2\sigma\cos(\alpha-2\pi/3)}
 \bigl[10\sigma\cos(\alpha-2\pi/3)-4\bigr],\\
 f_3^{(1)}&=\frac{\varepsilon_3}{3\sqrt3}\sqrt{1+2\sigma\cos(\alpha+2\pi/3)}
 \bigl[10\sigma\cos(\alpha+2\pi/3)-4\bigr],\\[4pt]
 f_1^{(2)}&=\frac{2\varepsilon_1\sigma}{3}\sqrt{1+2\sigma\cos\alpha}\sin\alpha,\\
 f_2^{(2)}&=\frac{2\varepsilon_2\sigma}{3}\sqrt{1+2\sigma\cos(\alpha-2\pi/3)}
 \sin(\alpha-2\pi/3),\\
 f_3^{(2)}&=\frac{2\varepsilon_3\sigma}{3}\sqrt{1+2\sigma\cos(\alpha+2\pi/3)}
 \sin(\alpha+2\pi/3).
 \end{aligned}}
\end{equation}
Here $\sigma=\sigma(\alpha,q)$.
All dependence on $q$ enters through the physical cubic root $\sigma(\alpha,q)$, except for the singlet, which already has the simple form in Eq.~\eqref{xyz-harmonic}.

In the original parameters, the theta-function representation is
\begin{equation}
 \begin{aligned}
 f_i^{(1)}&=\frac{\varepsilon_i\vartheta_4(t+\beta_i,q)}
 {R(q)\sqrt{\vartheta_4(t+\beta_j,q)\vartheta_4(t+\beta_k,q)}}
 \left[\frac{5\vartheta_4(t+\beta_i,q)^2}
 {R(q)^2\vartheta_4(t+\beta_j,q)\vartheta_4(t+\beta_k,q)}-3\right],\\[4pt]
 f_i^{(2)}&=\frac{\varepsilon_i
 \bigl[\vartheta_4(t+\beta_j,q)^3-\vartheta_4(t+\beta_k,q)^3\bigr]}
 {R(q)^3\bigl[\vartheta_4(t+\beta_j,q)\vartheta_4(t+\beta_k,q)\bigr]^{3/2}},
 \qquad (i,j,k)\text{ cyclic}.
 \end{aligned}
\end{equation}
For $0<q<1$, fractional powers are taken on the positive real branches of the theta factors, with the Cartesian signs carried by $\varepsilon_i$.

\subsection{Angular momentum, measure, and vibronic parity}
Each normalized basis function satisfies the spherical eigenvalue equation
\begin{equation}
 -\Delta_{S^2}\mathcal Y_{\ell a}=\ell(\ell+1)\mathcal Y_{\ell a}.
\end{equation}
The eigenvalue equation and orthonormality relations are preserved under reparametrization with the corresponding area measure: $d\Omega=J\,dt\,dq$ in the original map and $d\Omega=\mathcal J\,d\alpha\,dq$ in the orthogonal map, summed over the eight octants. In particular, the coordinate kinetic operator gives energies $6B_{\rm rot}$ for $\ell=2$ and $12B_{\rm rot}$ for $\ell=3$ in the absence of anisotropy. The cubic potential can mix different $\ell$ values, so these are kinetic eigenvalues, not general eigenvalues of the full Jahn--Teller Hamiltonian.

In O'Brien's reduced Jahn--Teller model, inversion parity determines the role of these functions. The $\ell=2$ functions are even under inversion and naturally describe the $E_g$ and $T_{2g}$ distortion coordinates and angular operators. They do not individually satisfy the antipodally odd nuclear-amplitude condition of the reduced model. The $\ell=3$ functions are odd and are admissible components of such nuclear amplitudes, alongside the $\ell=1$ triplet and higher odd degrees.

The explicit harmonic representations serve complementary purposes: the quadrupolar doublet identifies $\alpha$ and $\sigma$ directly with symmetry-adapted amplitudes; the cubic singlet is constant along each fixed-$q$ orbit; and the $T_{2u}$ triplet connects the harmonic basis to the Volterra evolution. The cubic singlet's $q$-only form does not imply separation of the full angular equation, since the metric coefficients and other harmonics retain angular dependence.

\paragraph{Algebraic verification.}
The polynomial representatives were checked symbolically to be harmonic and their full-sphere Gram matrices were evaluated exactly. Before normalization, the $\ell=2$ Gram matrix is $(4\pi/5)I_5$. For the ordered $\ell=3$ basis in Eq.~\eqref{octupole-basis}, it is diagonal with entries $4\pi/105$, three entries $16\pi/7$, and three entries $16\pi/105$. These exact integrals give the normalization factors stated above.

\section*{Discussion}
The theta-function map organizes the sphere into eight octants centered on trigonal directions. Bobek's cubic identity supplies its normalization, while an algebraic change of angle produces an orthogonal system with explicit metric and kinetic operator. The cubic-harmonic expressions give a symmetry interpretation of the new angle and connect the coordinate evolution to octupolar functions. In O'Brien's reduced Jahn--Teller model, these coordinates describe the geometry of trigonal wells and the associated vibrational and electronic quantities.

The formulas apply on the regular interiors of the coordinate patches. A global treatment of wavefunctions must retain smooth matching across octant boundaries, regularity at the centers, and the appropriate antipodal parity. The Volterra connection was identified with Mathematica and is supported by the high-precision checks reported in Section~4. The application establishes an explicit geometric reformulation of the reduced angular model, without assuming separation of its full eigenvalue equation.

\section*{Computational methods and AI assistance}
The development of the results presented here was enabled by combining AI-assisted mathematical analysis with symbolic computation in Wolfram Mathematica. The underlying theta identities, elliptic functions, and spherical harmonics are classical. Assembling these ingredients into explicit elliptic parametrizations of the $\ell=2$ and $\ell=3$ harmonics, together with the coordinate transformations and kinetic operators, requires nontrivial algebra, differentiation, branch choices, and normalization. A systematic derivation and verification by hand would be lengthy and laborious; the combined workflow made these calculations practical within the present study.

AI assistance was used to develop derivations, prepare computational code, organize the harmonic expressions, and refine the manuscript and illustrations. Mathematica was used for symbolic differentiation, exact polynomial and normalization checks, and high-precision numerical verification. Analytical arguments and numerical checks are identified separately in the text; the computational workflow supplements the mathematical justification of the formulas.

\end{document}